
\magnification=1200
{\nopagenumbers
\centerline{\bf Charged black holes in effective string theory}
\vskip40pt
\centerline{\bf S. Mignemi\footnote{$^\dagger$}{\rm Postal address: INFN,
Sezione di Cagliari,
via Ada Negri 18, 09127 Cagliari, Italy} and N. R. Stewart}
\vskip40pt
\centerline{Laboratoire de Physique Th\'eorique, Gravitation et Cosmologie
Relativiste}
\centerline{Universit\'e de Paris VI, CNRS/URA 769}
\centerline{Tour 22, 4e \'etage, B.C. 142  - 4, Place Jussieu,
75252 Paris Cedex 05, France}
\vskip40pt

{\noindent
We investigate the qualitative new features of charged dilatonic black holes
which emerge when both the Yang-Mills and Gauss-Bonnet curvature corrections
are included in the effective action. We consider these perturbative effects
by an expansion up to second order in the inverse string tension on the four
dimensional Schwarzschild background and determine the backreaction.
We calculate the thermodynamical functions and show that they can behave like
those of the Garfinkle-Horowitz-Strominger (GHS) solution [1] or in a more
conventional way, depending on the value of the magnetic charge $q$.
Moreover, we find that for magnetic charge above a critical value, the
temperature of the black hole has a
maximum and goes to zero for a finite value of the mass.
This indicates that the conventional Hawking evaporation law is modified by
string theory at a classical level.}
\vskip100pt
\centerline{November 1992}
\vfil\eject}

\def\lapp{[(r-2m)}
\def\alp{\alpha '}\def\GB{Gauss-Bonnet }\def\Sch{Schwarzschild }
\def\ri{{\cal R}}\def\gb{{\cal S}}\def\ef{{\rm e}^{-2\Phi}}
\def\fo{\phi_1}\def\ft{\phi_2}\def\ro{\rho_1}\def\rt{\rho_2}
\def\po{\psi_1}\def\pt{\psi_2}\def\en{{\rm e}^{2\nu(r)}}
\def\em{{\rm e}^{2\mu(r)}}
\def\lap{[r(r-2m)}\def\bc{boundary conditions }\def\af{asymptotically flat }

\noindent{\bf 1. Introduction}

It has been recently established [1-3] that the classical black hole
solutions of Einstein gravity in four dimensions
are endowed with new features when the theory is modified
by the introduction of the low energy string corrections to the action.
A key property of these modified
actions is the non-minimal coupling of the dilaton and axion fields with
gravity
and other fields. This fact permits to circumvent a class of well-known
"no-hair"theorems, which essentially state the triviality of non-minimally
coupled scalar fields [4].

In particular, if one neglects the Gauss-Bonnet corrections, exact charged,
spherically symmetric solutions with non trivial scalar hair are obtained in
ref. [1,5]. These have also been extended to the case of both dyonic and
axion charge in the spherically [3] and axially [6] symmetric case.
As is well known, if the mass of the black hole is large compared to the Planck
mass, then all vacuum solutions of Einstein gravity are approximate solutions
of low energy string gravity. The fact that the dilaton field couples to the
Yang-Mills field strength means that the charged black hole solutions in string
theory differ fundamentally from the classical Reissner-Nordstr\"om form.
Furthermore, the thermodynamical properties are quite unconventional: the
extremal dilaton black holes appear to have zero entropy but non-zero,
finite temperature [7]. This behaviour has been interpreted as due
to a repulsive
potential barrier created by the scalar field, which prevents thermic
contact with the exterior regions. It is then argued that these black holes
behave much like elementary particles [8]; important consequences ensue
for black hole evaporation [8,9].

Garfinkle et al. [1] obtained the exact classical solution for a four
dimensional charged black hole. They kept terms of order one in the inverse
string tension $\alp$ in the Yang-Mills field, but not in the curvature.
For large mass holes $(M\gg M_{Pl})$ one may truncate at this order and
neglect the same order curvature contributions outside the event horizon.
With an exact solution, comes the advantage of making statements about the
global properties of the solution. However, from the string theory point of
view the low energy effective action consists of an infinite perturbative
expansion in $\alp$. Of course near the singularity the curvature becomes so
large that all terms in the $\sigma$-model expansion would be important. But in
order to examine the structure of the event horizon, one may consider only
regions of finite curvature exterior to that region. Accordingly, in this
work we shall consider the perturbative effects up to second order in $\alp$
of both the Yang-Mills and curvature squared terms on the conventional
Schwarzschild  background solution. From this point of view, we are considering
the $O(\alp)$ string theoretic perturbations of Einstein gravity in vacuo.
In so doing, we shall determine the backreaction induced by these terms and the
dilaton on the geometry and the modified thermodynamics of the model.

We shall adopt units such that $\hbar=c=G=k=1$. So if the black hole is large
enough that $\alp\ll M^2$, then the curvature terms can be considered as
perturbations about the background solution. In particular we shall see that
the \GB terms contribute to the field equations for the metric only as
second order corrections whereas the electric or magnetic terms are $O(\alp)$
corrections to the vacuum \Sch background. Thus one can justify the neglect
of the \GB term as a first approximation far from the singularity as in [1].

Before proceeding with our analysis there are some remarks to be made
concerning
the form of the effective low energy string theory which we consider:

(i) The extra internal dimensions of heterotic string theory may be assumed to
decouple from the physical 4-dimensional spacetime by considering the topology
(4-dimensional spacetime)$\times$(internal compact space) and by imposing
appropriate boundary conditions on the internal coordinates. Alternatively,
as in the case of Calabi-Yau compactifications, one can "decouple" the
internal space by truncating all fields to their zero-modes on the internal
manifold. Thus we shall focus our attention on the four dimensional string
theory. Moreover, we shall not consider dilaton potentials which could be
produced by the effect of dimensional reduction.

(ii) The antisymmetric two-form $B_{mn}$, which arises in the ground state of
the string spectrum, appears in the action as the three-form axion field
strength $H_{mnp}$ coupled to the dilaton, where
$$H_{mnp}=\partial_{[m}B_{np]}+{\alp\over 4\sqrt 2}(\Omega_{mnp}^{(Y)}-
\Omega_{mnp}^{(L)})$$
$\Omega^{(L)}$ and $\Omega^{(Y)}$ are the Lorentz and Yang-Mills Chern-Simons
three-forms, respectively. Notwithstanding the Chern-Simons terms, the
non-minimal dilaton coupling to the axion field strength does not alter the
stationary spherically symmetric vacuum solutions of Einstein gravity (by a
duality transformation $H_{mnp}$ is equivalent to a pseudoscalar and so
the uniqueness theorem holds [10]). The situation is changed however for
axisymmetric spacetimes  where the Lorentz Chern-Simons term contributes to the
field equations and thus acts as a source for non trivial dilaton and axion
hair
[2,11]. Also in the case of dyons with both electric and magnetic charge,
the Yang-Mills Chern-Simons form acts as a source for $H_{mnp}$ through
the non-zero $F\wedge F$  term [12]. But for spherically symmetric
spacetimes, the Lorentz-Chern-Simons term can be expressed as the exterior
derivative of a two-form and thus absorbed into the definition of $B_{mn}$
[13]. In the present case we shall "freeze" in the $B_{mn}$ background
and discard its role. Since we aim to study the effect of the Gauss-Bonnet
term on the charged solutions of refs.[1,5], we shall not be considering the
dyonic case and thus the Yang-Mills Chern-Simons term vanishes in our case.
For these reasons we shall drop the axion field strength from the action,
consistently with the field equations.

The paper is organized as follows: in section 2 we introduce the action and
the field equation and discuss the limits of validity of our approximation.
In section 3 we discuss the case of zero
charge, which presents some interest by itself in the context of the "no-hair"
theorems of general relativity. In section 4 the perturbative expansion of the
GHS solution is studied in a gauge inspired by the exact solution
and compared with the exact result. Section 5 deals with
the full first order effective action including both the Maxwell and \GB terms:
the approximated solutions are obtained  in the same gauge and their
thermodynamical properties are discussed. Finally, in the appendix we rederive
the same results in a more conventional gauge.
\bigskip
{\noindent{\bf 2. The field equations}}

The bosonic sector of the effective action for the heterotic string is given
to leading order in $\alp$ as:
$$S_{eff}={1\over 16\pi}\int d^4 x \sqrt{-g}\left[\ri-2(\nabla\Phi)^2+\alpha
\ef(\gb-F^2)\right]
\eqno(1)$$
where $\alpha\equiv\alp /8$ and $\gb\equiv\ri^2_{abcd}-4\ri^2_{ab}+\ri^2$ is
the Gauss-Bonnet term. For the reasons given above, the terms containing the
axion have been omitted.

{}From variation of the action (1) we obtain the following equations of motion:
$$\eqalign{\ri_{mn}&=2\nabla_m\Phi\nabla_n\Phi
+2\alpha\ef (F_{mp}F_{np}-{1\over 4}
g_{mn}F^2)\cr
&+4\alpha\ef\Big[ 4\ri_{p(m}\nabla_{n)}\nabla_p\Phi-2\ri_{mn}\nabla_p\nabla_p
\Phi-g_{mn}\ri_{pq}\nabla_p\nabla_q\Phi\cr
&-\ri\nabla_m\nabla_n\Phi+{1\over2}
g_{mn}\ri\nabla_p\nabla_p\Phi-2\ri_{qmnp}\nabla_p\nabla_q\Phi\Big]\cr
&-8\alpha\ef\Big[ 4\ri_{p(m}\nabla_{n)}\Phi\nabla_p\Phi-2\ri_{mn}\nabla_p\Phi
\nabla_p\Phi-g_{mn}\ri_{pq}\nabla_p\Phi\nabla_q\Phi\cr
&-\ri\nabla_m\Phi\nabla_n
\Phi+{1\over2}
g_{mn}\ri\nabla_p\Phi\nabla_p\Phi-2\ri_{qmnp}\nabla_p\Phi\nabla_q\Phi\Big]\cr}
\eqno(2a)$$
$$\nabla^2\Phi={\alpha\over 2}\ef (\gb -F^2)\eqno(2b)$$
$$\nabla_p(\ef F_{pm})=0\eqno(2c)$$

In order to find approximate solutions to (2) we shall expand the fields
around the background constituted by the Schwarzschild metric with vanishing
dilaton, which is of course a solution of the field equations. Our expansion
will be in the parameter $\alpha$ or, more correctly, in ${\alpha\over m^2}$,
$m$ being the mass of the \Sch solution.
Since $\alpha$ is believed to be of order unity in Planck units, the
expansion
is valid for large $m$, in the regions where $\alpha\gb\ll\ri$,
i. e. for $r^3\gg\alpha m$.
For macroscopic black holes ($m\gg 1$) this condition is always satisfied,
except in a neighborood
of the singularity, well inside the horizon (region of strong curvature).
Near the physical singularity, however, the higher order corrections to the
effective string lagrangian become important and so the perturbation theory
is no longer reliable.
For charged holes one must also require that $\alpha F^2\ll\ri$.
This condition holds
for $r\gg{\alpha q^2\over m}$; again, if the charge is not too large, this
condition is satisfied by macroscopic black holes far from the singularity.

Alternatively, one might study the effect of the \GB term by perturbing the
exact charged GHS solution. The perturbation would be in the parameter ${q^2
\over m}$, and would be valid for $F^2\gg\gb$, i.e. $r\gg m$. The range of
validity of this alternative expansion is therefore smaller. Moreover, the
physical interpretation as expansion in the string
parameter $\alpha '$ would be lost. For these reasons we will not adopt
this point of view.
\bigskip
\noindent{\bf 3. The q=0 case}

In this section we study the pure graviton-dilaton system, by making the
ansatz $F=0$, consistently with the equations of motion. Besides its interest
in connection with string theory, this special case is interesting
because it offers an
example of non-trivial scalar hair in a purely scalar-gravity theory. The
relevance of the Gauss-Bonnet term for gravity is due to the fact that it is
the only quadratic lagrangian which gives rise to second order field equations
for the metric [14]. Moreover it does not introduce new propagating degrees of
freedom into the ghost-free theory [15].
For constant $\Phi$ however, this term does not contribute to the field
equations, since it is a total derivative in four dimensions. Only its
coupling with the dilaton can therefore give rise to non-trivial solutions in
four dimensions.

Similar models have been studied in refs.[16-19]. Ref. [16] and [17],
however, consider the higher-dimensional case but do not treat in detail the
4-dimensional limit, where the first order corrections due to the \GB term
vanish. In ref. [17] is shown that, in dimensions higher than four, at
first order in $\alp$, the temperature has a maximum and then goes to zero
for a finite value of the mass of the hole\footnote\dag{ Actually, in ref.
[17] the \GB term is replaced by the square of the Riemann tensor. At first
order of
an expansion around the \Sch background the results are however the same,
since $\bar\ri_{mn}=\bar\ri=0$ for the \Sch solution.}.
This would mean that a stable
configuration could be reached during the evaporation of the black hole.
However, as we shall see, this is not the case in four dimensions, because
then the correction to the temperature has the opposite sign.

The four dimensional case has been discussed in refs. [18,19] in connection
with Kaluza-Klein theories; there the relevance of the model for the
no-hair theorems was pointed out. However, an essential difference from our
treatment is
the fact that Poisson [19] does not perform the conformal rescaling which puts
the lagrangian in its canonical form. This canonical form should be used
for reasons of stability [20] and because it gives an anambiguous
definition of the physical mass [17].

We shall look for the spherical symmetric solutions of the field equations (2)
by perturbing around the
background given by the \Sch solution with vanishing scalar field ($\phi_0=0$).
For vanishing $F$ the scalar field equation reduces to:
$$\nabla^2\Phi={\alpha\over 2}\ef\gb
\eqno(3)$$
where the \GB term $\gb={48m^2\over r^6}$ for the Schwarzschild
background. Expanding $\Phi$ as
$$\Phi(r)=\alpha\fo(r)+\alpha^2\ft(r)+\dots$$
one has at first order
$$\lap\fo ']'={24m^2\over r^4}
\eqno(4)$$
where the prime denotes the derivative with respect to $r$.

The only solution of (4) regular at the horizon $r=2m$, is given by (apart
from an arbitrary constant which we fix to zero, by requiring $\fo\to 0$ at
infinity):
$$\fo =-{1\over m}\left( {1\over r}+{m\over r^2}+{4m^2\over 3r^3}\right)
\eqno(5)$$
This solution has a dilatonic charge ${\alpha\over m}$, corresponding to a
long range repulsive potential. As is usual models of this type, this charge
is not a new independent parameter [1].

One can now calculate the backreaction on the metric which is induced by the
scalar field, by substituting
(5) into the gravitational equations
(2a). It is easy to see that, contrary to the higher dimensional case [17]
the first contribution to the r.h.s. arises only at order $\alpha^2$.
In fact, in four dimensions,
at order $\alpha^2$, the r.h.s. of (2a) reduces to
$$2\nabla_m\fo\nabla_n\fo-8\bar\ri_{bmna}\nabla_a\nabla_b\fo
\eqno(6)$$
where $\bar\ri_{bmna}$ is the value of the Riemann tensor in the \Sch
background.

The static, spherically symmetric metric can now be written as
$$ds^2=-\en dt^2+\em dr^2+r^2 d\Omega^2
\eqno(7)$$
and the metric functions can be expanded as
$$\nu(r)=\nu_0(r)+\alpha^2\tau(r)\qquad\qquad\mu(r)=\mu_0(r)+\alpha^2\sigma
(r)
\eqno(8)$$
with ${\rm e}^{2\nu_0}={\rm e}^{-2\mu_0}=1-{2m\over r}$, corresponding to the
\Sch solution; we have taken into account that there are no corrections to the
metric functions at
order $\alpha$.
Substituting for the metric functions $\mu(r)$ and $\nu(r)$ in eqn.(2a), one
finds two independent
field equations:
$$\sigma '+\tau '=r\fo '^2-{8m\over r^2}\fo''
\eqno(9a)$$
$$\sigma '-\tau '+{2\sigma\over r-2m}=-{8m\over r^2}\fo ''+
{16m(r-3m)\over r^3(r-2m)}\fo '
\eqno(9b)$$
and, upon combining the two,
$$[(r-2m)\sigma]'={1\over 2}r(r-2m)\fo '^2+8m\left[{r-3m\over r^3}\fo '-
{r-2m\over r^2}\fo ''\right]
\eqno(10)$$
We notice from eqn. (9a) that in presence of a scalar field the metric
functions
$\mu$ and $-\nu$ are not equal.

Substituting (5) in (10) and integrating, one obtains for $\sigma$:
$$\sigma={1\over 2m^4}\left({49m\over 40r}+{29m^2\over 20r^2}+{19m^3\over
10r^3}
-{203m^4\over 15r^4}-{436m^5\over 15r^5}-{184m^6\over 3r^6}\right)
\eqno(11)$$
where the integration constant has been chosen such that $\sigma$ is regular
at the horizon $r=2m$. The expression for $\tau$ can now be immediately
obtained from (9a):
$$\tau=-{1\over 2m^4}\left({49m\over 40r}+{49m^2\over 20r^2}+{137m^3\over
30r^3}
+{7m^4\over 15r^4}-{52m^5\over 15r^5}-{40m^6\over 3r^6}\right)
\eqno(12)$$
with the same boundary conditions as for $\sigma$.

The gravitational and inertial mass $m_G$ and $m_I$ are defined respectively as
$$-g_{00}\approx 1-{2m_G\over r}+O\left({1\over r^2}\right)\qquad\qquad
g_{11}\approx 1+
{2m_I\over r}
+O\left({1\over r^2}\right).
\eqno(13)$$
The two values coincide and are given by
$$M\equiv m_G=m_I=m\left(1+{49\alpha^2\over 80m^4}\right).
\eqno(14)$$
The mass of the black hole is therefore increased with respect to that of
the \Sch
solution, or better, the horizon is shrunk to $r_0=2M(1-49\alpha^2/80M^4)$.

It is now easy to calculate the temperature $T$ of the hole, by requiring the
regularity of the euclidean section of the metric at the horizon [21]. The
result is
$$\beta\equiv{1\over T}=8\pi m[1+\alpha^2(\sigma -\tau)]\Bigl\arrowvert_{r=2m}=
8\pi m\left[1+{\alpha^2\over 240m^4}\right]
\eqno(15)$$
Using (14) and (15) the temperature can be written in terms of the physical
mass $M$:
$$T={1\over 8\pi M}\left(1+{73\alpha^2\over 120M^4}\right),
\eqno(16)$$
which is higher than that of the \Sch hole of equal mass.

Contrary to what happens in higher dimensions [17], the temperature is a
monotonically decreasing function of the mass, and therefore no mechanism
to prevent the complete evaporation of the black hole seems to be available in
this case. As we shall see, the situation may change however for
charged black holes.

The entropy is easily obtained from its definition in terms of the euclidean
action (see section 4), and is given by:
$$S=4\pi M^2\left(1+{\alpha\over M^2}+{73\alpha^2\over120M^4}\right).
\eqno(17)$$
As is usual in higher derivative theories [22], it does not coincide
with the area of the event horizon. It is, however, a positive definite
function of M.
\bigskip
\noindent{\bf 4. The purely magnetic case in the GHS gauge}

In order to study the effect of the Maxwell field it can be useful to write
the metric in a different form. The reason is that an exact solution of the
Einstein-Maxwell-dilaton system is known in the absence of the Gauss-Bonnet
term [1,5], which has a very simple expression in a different "gauge" (i.e.
coordinate system). For a magnetically charged hole it can be written as [1]:
$$ds^2=-\lambda^2 dt^2+\lambda^{-2}dr^2+R^2d\Omega^2
\eqno(18)$$
where
$$\lambda^2=1-{2m\over r}\qquad\qquad R^2=r\left(r-{\alpha q^2\over m}\right)
\eqno(19a)$$
and
$$F=q\sin\theta d\theta\wedge d\varphi\qquad\qquad\ef=1-{\alpha q^2\over mr}
\eqno(19b)$$
where $q$ is the magnetic charge ($q$ is related to the physical charge $Q$ by
$Q=\sqrt{\alpha/4\pi}q$).
We start by discussing the expansion for the case in which the
Maxwell term is present in the lagrangian, while the \GB term is absent. In
this
way we shall compare the perturbative solution with the exact one and acquire
some confidence with the GHS gauge.

We therefore write the metric in the form (19) and expand the metric functions
around a \Sch background as:
$$\lambda=\lambda_0(1+\alpha\po +\alpha^2\pt+\dots)\ ;\qquad\qquad
R=r+\alpha\ro+\alpha^2\rt+\dots
\eqno(20a)$$
$$\Phi=\alpha\fo+\alpha^2\ft+\dots ,
\eqno(20b)$$
where $\lambda_0=(1-{2m\over r})^{1/2}$ and $\psi_i$, $\rho_i$ and $\phi_i$
are functions of $r$.

The equation (2c) is solved by the ansatz
$$F_{ij}={q\over R^2}\epsilon_{ij}
\eqno(21)$$
The equation for the dilaton becomes at first order in $\alpha$:
$$[r(r-2m)\fo ']'=-{q^2\over r^2}
\eqno(22a)$$
which, up to an integration constant yields
$\fo={q^2\over 2mr}$.
The other equations are given at order $\alpha$ by
$$\ro ''=0
\eqno(22b)$$
$$\lapp\po]'=-{m\over r^2}\ro-{q^2\over 2r^2}
\eqno(22c)$$
We impose the boundary conditions that $\ro\to$const, $\po\to 0$ at
infinity. We are still free to choose the boundary conditions at $r=2m$.
Changing the \bc at $r=2m$ yields a reparametrization of the solutions, but no
change in their physical properties: in particular, the relations between the
physical quantities, like mass, temperature and entropy, are independent of
the parametrization.
The simplest choice for the following calculations is to choose
for $\ro$, which must be constant in view of (22b) and the \bc at infinity,
the value
$\ro=-{q^2\over 2m}$, so that the r.h.s. of (22c) vanishes. This choice
permits us to have
$\po=0$.

We can now evaluate the second order corrections. For vanishing $\po$ the
equations are given by:
$$\lap\ft']'=2{r-2m\over r}\ro\fo '+{2q^2\over r^2}\left(\fo+{2\ro\over r}
\right)={q^4\over 2m^2}\left({1
\over r^2}-{4m\over r^3}\right)
\eqno(23a)$$
$$\rt''=-r\fo'^2=-{q^4\over 4m^2r^3}
\eqno(23b)$$
$$\lapp\pt]'=-{r-2m\over 2}\rt ''-{r-m\over r}\rt '-{m\over r^2}\rt
+{2m\over r^3}\ro^2+{q^2\over r^2}\left(\fo+{2\ro\over r}\right)=0
\eqno(23c)$$
The solutions satisfying the \bc stated above are easily seen to be:
$$\ft={q^4\over 4m^2r^2}\ ,\qquad\rt=-{q^4\over 8m^2r}\ ,\qquad\pt=0
\eqno(24)$$

It is now straightforward to compare the expansion with the exact result (19):
it turns
out that it coincides with the expansion of the exact solution in powers of
$\alpha$:
$$\Phi=-{1\over 2}\ln\left(1-{\alpha q^2\over mr}\right)\sim{\alpha q^2
\over 2mr}+{\alpha^2 q^4\over 4m^2r^2}
\eqno(25a)$$
$$\lambda=\lambda_0=\sqrt{1-{2m\over r}}
\eqno(25b)$$
$$R=\sqrt{r\left(r-{\alpha q^2\over m}\right)}\sim r-{\alpha q^2\over 2m}
-{\alpha^2q^4\over8m^2r}
\eqno(25c)$$
The physical properties of this solution are discussed in [1,8] and we
shall not report them here. It may be
worth noticing, however, that the values of the temperature and entropy of
the hole in this approximation, obtained up to second order in $\alpha$ with
the
methods described in the next section, turn out to coincide with the exact
result [8]:
$$T={1\over 8\pi M}\ ,\qquad\qquad S=4\pi\left(M^2-{\alpha q^2\over 2}\right)
\eqno(26)$$
where $M=m$ is the physical mass, as is evident from (25b).
\bigskip

\noindent{\bf 5. Solutions of the full order $\alpha '$ effective action}

We now consider the more interesting case where both the Maxwell and
Gauss-Bonnet terms are present.
As we shall see, whereas the Maxwell term contributes already at first order
to the metric functions, the \GB terms contributes only at second order.
This fact gives some justification to the interpretation of the GHS
solution as a
first order approximation for the metric. However, this is not true for the
scalar field, to
which the \GB terms contribute already at first order. With the ansatz (21)
for $F$, in
the notation of the previous section, the  field equations (2) become at
order $\alpha$:
$$\lap\fo ']'=-{q^2\over r^2}+{24m^2\over r^4}
\eqno(27a)$$
$$\ro ''=0
\eqno(27b)$$
$$\lapp\po]'=-{m\over r^2}\ro-{q^2\over 2r^2}
\eqno(27c)$$
Again, we impose \bc such that the metric is \af and exploit the remaining
freedom to fix $\po =0$. Integration of (27) yields:
$$\fo=-{1\over m}\left({2-q^2\over 2r}+{m\over r^2}+{4m^2\over 3r^3}\right)
\eqno(28a)$$
$$\ro=-{q^2\over 2m}\ ,\qquad\qquad\po=0
\eqno(28b)$$
The dilatonic charge is now given at first order by ${2-q^2\over 2m}\alpha$.
The long range
potential driven by the scalar field can therefore be attractive or repulsive
depending on the value of
$q^2$.

We go now to second order in $\alpha$. The field equations become:

$$\eqalign{\lap\ft']'&=2{r-2m\over r}\ro\fo '+{2q^2\over r^2}
\left(\fo+2{\ro\over r}
\right)-{48m^2\over r^4}\left(\fo+{\ro\over r}\right)\cr
&={2q^2-q^4\over 2m^2r^2}+2{q^2+q^4\over m r^3}+{2q^2\over r^4}+16m{9+q^2\over
3r^5}+{48m^2\over r^6}+{64m^3\over r^7}}
\eqno(29a)$$
$$\eqalign{\rt''&=-r\fo'^2+{8m\over r^2}\fo''\cr
&=-\left[{(q^2-2)^2\over 4m^2r^3}+2{2-q^2\over mr^4}+4{7-3q^2\over r^5}
+{64m\over r^6}+{144m^2\over r^7}\right]}
\eqno(29b)$$
$$\eqalign{\lapp\pt]'&=-{r-2m\over 2}\rt ''-{r-m\over r}\rt '
-{m\over r^2}\rt
+{2m\over r^3}\ro^2\cr\quad&+{q^2\over r^2}\left(\fo+{2\ro\over r}\right)
+4m\left(
{r-2m\over r^2}\fo''-2{r-3m\over r^3}\fo'\right)\cr
&=2{1-2q^2\over 3mr^3}-{11-5q^2\over r^4}+8m{1-25q^2\over 15r^5}
+{544m^3\over 5r^7}}
\eqno(29c)$$

It is no longer possible to have a vanishing $\pt$. We therefore choose the
boundary conditions such that $\lambda^2\approx 1-{2m\over r}+O\left({1\over
r^2}
\right)$ at infinity, so that $m$ is by definition the physical mass $M$ of
the hole.
The solutions of (29) are:
$$\ft=-\left[{73-45q^2\over 60m^3r}+{73-15q^2-15q^4\over 60m^2r^2}+{73\over
45mr^3}+{73+5q^2\over 30r^4}+{112m\over 75r^5}+{8m^2\over 9r^6}\right]
\eqno(30a)$$
$$\rt=-\left[{(q^2-2)^2\over 8m^2r}+{2-q^2\over 3mr^2}+{7-3q^2\over 3r^3}+
{16m\over 5r^4}+{24m^2\over 5r^5}\right]
\eqno(30b)$$
$$\tilde\pt\equiv (r-2m)\pt=-\left[{1-2q^2\over 3mr^2}-{11-5q^2\over 3r^3}
+{(2-50q^2)m\over
15r^4}+{272m^3\over 15r^6}\right]
\eqno(30c)$$

The metric functions are therefore:
$$\eqalign{R^2&=r^2+2\alpha r\ro+\alpha^2(2r\rt+\ro^2)\cr
&=r^2-{\alpha q^2\over m}r-
\alpha^2\left[{1-q^2\over m^2}+2{2-q^2\over 3mr}+2{7-3q^2\over 3r^2}+
{32m\over 5r^3}+{48m^2\over 5r^4}\right]}
\eqno(31a)$$
$$\eqalign{\lambda^2&=1-{2m\over r}+2\alpha^2{\tilde\pt\over r}\cr
&=1-{2m\over r}-2\alpha^2\left[{1-2q^2\over3mr^3}-{11-5q^2\over 3r^4}
+{(2-50q^2)m\over
15r^5}+{272m^3\over 15r^7}\right]}
\eqno(31b)$$
They are sketched in Figure 1, for different values of $q$.
For $q=0$ one obtains of course the solution discussed in section 2, but in
different coordinates. In particular, in the present coordinates the horizon is
at $r_+=2m(1-\alpha^2\tilde\pt(2m))$ $=2m\left(1-{1+2q^2\over12m^4}\alpha^2
\right)$. We notice that the coordinate $r$ should not be identified with
the physical radial coordinate, which is rather given by $R$. Actually, $r$
must
vary in the range $r\ge r_-$, where $r_-\approx{\alpha q^2\over m}$ is the
greatest zero of $R^2$, which corresponds to a singular surface.
When $r_-$ becomes greater than $r_+$ (for $\alpha q^2>2m^2(1+7\alpha/12)$),
one has a naked singularity. This
situation is, however, beyond the range of validity of our approximation.

It is now easy to calculate the temperature of the hole. The inverse
temperature
is defined as the periodicity of the time coordinate which renders
the euclidean section of the solution regular and is given by [21]:
$$\beta=4\pi\sqrt{g_{00}g_{11}}\left[{d\over dt}g_{00}\right]^{-1}\Bigl
\arrowvert_{r=2m}
\eqno(32)$$
In our coordinates eqn.(32) becomes at order $\alpha^2$:
$$\beta=8\pi m \left[1-\alpha^2\left({\tilde\pt(2m)\over m}-2\tilde\pt '(2m)
\right)\right]=8\pi m \left(1-\alpha^2{73-45q^2\over 120m^4}\right)
\eqno(33)$$
It is interesting to notice that the temperature of the black hole, which is
given by
$$T={1\over8\pi M}\left(1+\alpha^2{73-45q^2\over 120M^4}\right)
\eqno(34)$$
has no order $\alpha$ corrections (we recall that the physical mass $M=m$ in
these coordinates).
The behaviour of $T$ as a function of $M$
is displayed in Figure 2a for several values of $q$.

It is remarkable the fact that the temperature is no longer independent of
the charge, as in the GHS solution, but has a very different behaviour
depending
on the values of $q$. In particular, if $q^2>73/45$, the temperature vanishes
for $M^4=\alpha^2({3\over 8}q^2-{73\over
120})$. The hole may
therefore reach during its evaporation a stable groundstate with nonvanishing
values of mass and charge\footnote\dag{The implications for cosmology of a
model with similar behaviour of $T$ have been recently discussed in
[23] and, in their notation, $m^4_{rel}\propto({3q^2\over 8}-{73\over 120})
m^4_{Pl}$.}.
Unfortunately, however, for such values of $q$ and $M$ our
approximation breaks down and therefore it is not possible to discuss the
causal
structure of the solution. In particular, we are not able to establish if the
states with vanishing temperature correspond to regular horizons or naked
singularities.

The entropy of the hole can be calculated by evaluating the euclidean
action [24]:
$$I_E=-{1\over 16\pi}\int_M[\ri-2(\nabla\Phi)^2+\alpha\ \ef(\gb-F^2)]dV-
{1\over 8\pi}
\int_{\partial M}(K-K_0+{\cal Q})d\Sigma
\eqno(35)$$
where the boundary integral includes a term ${\cal Q}$ coming from the \GB
part of the action.
This term, which is proportional to the Chern-Simons form on the boundary
[22], gives vanishing contribution in four dimensions.

By using the equations of motion (2) and neglecting $O(\alpha^3)$ terms,
$I_E$ can therefore be reduced to the form:
$$\eqalign{I_E&=-{\alpha\over 16\pi}\int_M\ef(\gb-F^2)-{1\over 8\pi}\int_{
\partial M}(K-K_0)=
-{1\over 8\pi}\int_M\nabla^2\Phi-{1\over 8\pi}\int_{\partial M}(K-K_0)\cr
&={1\over 2}\beta m\left( 1-\alpha {2-q^2\over 2m^2}-
\alpha^2{73-45q^2\over 60 m^4}\right)=4\pi m^2\left(1-\alpha{2-q^2\over
2m^2}-\alpha^2{73-45q^2\over 40m^4}\right)}
\eqno(36)$$

The mass $M$ of the hole can be defined as $M={\partial I\over\partial
\beta}|_q$, where
$${\partial\over\partial\beta}\Big\arrowvert_q={\partial m\over\partial\beta}
{\partial\over
\partial m}={1\over 8\pi}\left(1-\alpha^2{73-45q^2\over 40m^4}\right)
{\partial\over\partial m}
\eqno(37)$$
One has therefore, by (36),
$$M={1\over 2}\left(1-\alpha^2{73-45q^2\over 40m^4}\right)
{\partial\over\partial m}\left(1-\alpha^2{73-45q^2\over 40m^4}\right)=m,
\eqno(38)$$
as expected. The entropy is then given by:
$$S=\beta{\partial I\over\partial\beta}-I=4\pi M^2\
\left( 1+\alpha{2-q^2\over 2M^2}+
\alpha^2{73-45q^2\over 120 M^4}\right)
\eqno(39)$$
Its behaviour as a function of $M$ is displayed in figure 2b.

We note that the entropy has a zero for $\alpha q^2=2M^2(1+{\alpha
\over 4M^2}
+{101\alpha^2\over 240M^4})$. For the exact GHS solution the zero of the
entropy
corresponds to the case of extremal black holes [8]. By comparing (34) and (39)
it is easy to see that, for positive $\alpha$, if $q^2>2$ the zero of the
entropy occurs for values of $M$ greater than those for which $T=0$, while for
$q^2<2$ one has the opposite situation. It seems therefore that when the \GB
terms are taken into account, for small or zero charge the situation is more
similar in this respect to the usual black hole thermodynamics, while for
greater charge one
has physical properties more similar to those of the GHS solution [8].
It may be
useful to remark that this critical value of $q^2$ is precisely that for which
the long range scalar potential change from attractive to repulsive at first
order in $\alpha$ (see eq.(28a)).

Of course, we are considering only the lower order corrections to $T$ and $S$.
When higher terms of the expansion are taken into account the situation may
change. In particular, for small values of $M$ our approximation is not
reliable.
Our results should be essentially unchanged, however, for large values
of mass and charge.
\bigskip
\noindent{\bf 6. Final remarks}

We have studied the effect of the inclusion of the \GB corrections
induced by string theory on the spherically symmetric
solutions representing the charged dilatonic black holes obtained in [1,5].
The physical properties of the solution resemble
those of the GHS solution for large charge. For small or vanishing charge
the situation looks similar to the more conventional theories.

In particular, the \GB term induces $O(\alpha^2)$ charge-dependent corrections
to the temperature. Moreover, for $q^2>{73\over 45}$, our stringy black hole
has vanishing temperature for a finite positive value of $M$. Our analysis
thus
illustrates the importance of the higher order curvature correction for the
thermodynamics of an evaporating classical black hole.

We have considered the string effective action up to order $\alpha '$. The
results obtained are not essentially affected if one takes into account the
order $\alpha '^2$ corrections to the effective action, as given in [26].
The new terms, in fact, give contributions of order $\alpha '^2$ to $\Phi$
and of order $\alpha '^3$ to the metric fields. The formula (34) for the
temperature,
in particular, is corrected only by third order terms.

Our results may be extended to the case of electric charge. Unfortunately,
due to the presence of the \GB term, the electromagnetic duality which allows
one to obtain immediately the electrically charged solution from the magnetic
one [3] is no longer a symmetry of the field equations.
The calculation should therefore be started ab initio. Analogous considerations
hold for the dyonic case, with the further complication that the axion
field should also be taken into account in this case [13]. These problems are
currently under investigation.

Other interesting perspectives are given by the study of massive dilatons. The
dilaton potential generated by the compactification mechanism may induce some
modifications in the properties of the black holes.
Some results have been obtained in ref. [27], in the absence of the \GB term.
\bigskip
\noindent{\bf Acknowledgements}

N.R.S. wishes to thank the Royal Society for a European fellowship. The work of
S.M. was supported by a MRST fellowship.
\bigskip
\def\so{\sigma_1}\def\st{\sigma_2}\def\to{\tau_1}\def\tt{\tau_2}
\noindent{\bf Appendix}

In this appendix we rederive the results of section 5 in a more conventional
gauge. We adopt the metric form (7) and expand the metric functions as:
$$\nu=\nu_0+\alpha\to+\alpha^2\tt\dots\quad\qquad\mu=\mu_0+\alpha\so+\alpha^2
\st
\eqno(A1)$$
where $\nu_0$ and $\mu_0$ are the values of $\mu$ and $\nu$ in the \Sch
background of mass $m$. Due to the presence of the Maxwell field, order
$\alpha$
corrections to the metric functions must also be taken into account.
The parameter
$m$ and the radial coordinate $r$ used in this appendix do not coincide of
course with those of section 5. The relations between the physical quantities
(mass, temperature, entropy...) are nevertheless the same in the two gauges.

With the ansatz (21) for $F$, the field equation  (27a) for $\fo$ is
unchanged in this gauge, and the solution is still given by (28a).

The gravitational equations become at first order:
$$\to '+\so '=0
\eqno(A.2a)$$
$$(r-2m)(\so '-\to ')+2\so={q^2\over r^2}
\eqno(A.2b)$$
By requiring asymptotic flatness and regularity at the horizon, the solutions
of (A.2) are given uniquely by:
$$\so=-\to={q^2\over 4mr}
\eqno(A.3)$$
Using (A.3) the field equation for the second order corrections to the scalar
field can be written as:
$$\lap\ft ']'={(r-2m)q^2\over 2mr^2}\fo ''+{q^2\over 2mr^2}\fo '+{2\over r^4}
\left[q^2-{48m^2\over r^2}\right]\fo+{12q^2\over r^6}\left[1-{4m\over r}\right]
\eqno(A.4)$$
which has solution:
$$\eqalign{\ft=&-{1\over 360m^4}\Big[(438-360q^2+45q^4){m\over r}+(438-270q^2)
{m^2\over
r^2}+(548-240q^2){m^3\over r^3}\cr&+(876-660q^2){m^4\over r^4}+{5376\over 10}
{m^5\over r^5}
+320{m^6\over r^6}\Big]\cr}
\eqno(A.5)$$
where as usual we require regularity at the horizon and asymptotic flatness.
The gravitational equations are at second order:
$$\tt '+\st '=r\fo '^2-{8m\over r^2}\fo ''
\eqno(A.6a)$$
$$\st '-\tt '+{2\st\over r-2m}=-{8m\over r^2}\fo ''+{16m(r-3m)\over r^3(r-2m)}
\fo '-{2q^2\over r^2(r-2m)}\fo -{(r-4m)q^4\over 8m^2r^3(r-2m)}
\eqno(A.6b)$$
With the usual boundary conditions their solutions are:
$$\eqalign{\tt=&-{1\over 480m^4}\Big[(294-30q^2-15q^4){m\over r}+(588-60q^2)
{m^2\over
r^2}+(1096-200q^2){m^3\over r^3}\cr&+(112+560q^2){m^4\over r^4}-
832{m^5\over r^5}
-3200{m^6\over r^6}\Big]\cr}
\eqno(A.7a)$$
$$\eqalign{\st=&{1\over 480m^4}\Big[(294-30q^2-15q^4){m\over r}
+(348+180q^2-60q^4)
{m^2\over
r^2}+(456+120q^2){m^3\over r^3}\cr&-(3248-2000q^2){m^4\over r^4}-6976
{m^5\over r^5}
-14720{m^6\over r^6}\Big]\cr}
\eqno(A.7b)$$
The  physical mass $M$ of the hole, defined as in (13) is given by:
$$M\equiv m_G=m_I=m\left[1+\alpha{q^2\over 4m^2}+\alpha^2{294-30q^2-15q^4\over
480m^4}\right]
\eqno(A.8)$$
Inverting, one can write the parameter $m$ as a function of the physical mass:
$$m=M\left[1-\alpha{q^2\over 4M^2}-\alpha^2{294-30q^2+15q^4\over 480M^4}\right]
\eqno(A.9)$$
One can now deduce the temperature from (32): one has, up to second order:
$$\eqalign{\beta^{-1}\equiv T&={1\over 8\pi m}[1+\alpha(\to-\so)+\alpha^2
(\tt-\st)+
{\alpha^2\over 2}(\to-\so)^2]\Big\arrowvert_{r=2m}\cr
&={1\over 8\pi m}\left[1-\alpha{q^2\over 4m^2}-\alpha^2{2+150q^2-45q^4\over
480m^4}\right]\cr}
\eqno(A.10)$$
Using (A.9) one recovers the result of section 5 for the temperature in terms
of the physical mass:
$$T={1\over 8\pi M}\left[1+\alpha^2{73-45q^2\over 120M^4}\right]
\eqno(A.11)$$
Analogously, following the procedure of section 5, one can evaluate the
entropy;
the euclidean action turns out to be:
$$\eqalign{I_E&={\beta\over 2}\left[m-\alpha{4-3q^2\over 4m}-\alpha^2
{290-450q^2+75q^4\over
480m^3}\right]\cr&=4\pi M^2\left[1-\alpha{2-q^2\over2M^2}-\alpha^2
{73-45q^2\over 40M^4}\right]\cr}
\eqno(A.12)$$
which is the result of section 5. The value of the entropy is therefore given
by (39).
\bigskip

\def\PRD{Phys. Rev. D}\def\NPB{Nucl. Phys. B}\def\PL{Phys. Lett. }
\def\MPLA{Mod. Phys. Lett. A}\def\CMP{Comm. Math. Phys. }
\def\PRL{Phys. Rev. Lett. }\def\JMP{J. Math. Phys. }
\def\PTP {Prog. Theor. Phys. }
\def\CQG{Class. Quantum Grav. }
\centerline{\bf References}
\halign{#\hfil&\ #\hfil\cr
[1]& D. Garfinkle, G.T. Horowitz, A. Strominger, \PRD 43, 3140 (1991)\cr
[2]& B.A. Campbell, M.J. Duncan, N. Kaloper, K.A. Olive, \PL 251B, 34 (1990)\cr
[3]& A. Shapere, S. Trivedi, F. Wilczek, \MPLA 6, 2677 (1991) \cr
[4]& J.D. Bekenstein, \PRD 5,1239 (1972)\cr
& J.E. Chase, \CMP 19, 276 (1970)\cr
[5]& G.W. Gibbons, K. Maeda, \NPB 298, 741 (1988)\cr
[6]& A. Sen, preprint TIFR-TH-92-90, TIFR-TH-92-41 (1992)\cr
[7]& J. Preskill, P. Schwarz, A. Shapere, S. Trivedi, F. Wilczek, \MPLA 6, 2353
\cr&(1991)\cr
[8]& C.F.E. Holzhey, F. Wilczek, \NPB 380, 447 (1992)\cr
[9]& S.B. Giddings, A. Strominger, \PRD 46, 627 (1992)\cr
[10]& M.J. Bowick, S.B. Giddings, J.A. Harvey, G.T. Horowitz, A. Strominger,
\PRL\cr&61, 2823 (1988)\cr
[11]& S. Mignemi, N.R. Stewart, preprint, in press, \PL B (1992)\cr
[12]& B.A. Campbell, M.J. Duncan, N. Kaloper, K.A. Olive, \PL 263B, 364 (1990)
\cr
& B.A. Campbell, N. Kaloper, K.A. Olive, \PL 285B, 199 (1990)\cr
& K. Lee, E.J. Weinberg, \PRD 44, 3159 (1991)\cr
[13]& B.A. Campbell, M.J. Duncan, N. Kaloper, K.A. Olive, \NPB 351,778
(1991)\cr
[14]& D. Lovelock, \JMP 12, 498 (1991)\cr
[15]& B. Zwiebach, \PL 165B, 315 (1985)\cr
[16]& D.G. Boulware, S. Deser, \PL 175B, 409 (1986)\cr
[17]& C.G. Callan, R.C. Myers, M.J. Perry, \NPB 311, 673 (1988)\cr
[18]& A. Tomimatsu, \PTP 79, 86 (1988)\cr
[19]& E. Poisson, \CQG 8, 639 (1991)\cr
[20]& L. Sokolowski, \CQG 6, 2045 (1989)\cr
[21]& J.W. York, \PRD 31, 775 (1985)\cr
[22]& B. Whitt, \PRD 38, 3000 (1988)\cr
&R.C. Myers, J.Z. Simon, \PRD 38, 2434 (1988)\cr
[23]& J.D. Barrow, E.J. Copeland, A.R. Liddle, preprint Sussex-AST-92/3-1
(1992)\cr
[24]& S.W. Hawking, in {\it "General Relativity: an Einstein centenary
survey"}, eds. S.W. Hawking\cr&and W. Israel (Cambridge Un. Press 1979)\cr
[25]& R.C. Myers, \PRD 36, 392 (1987)\cr
&F. M\"uller-Hoissen, \NPB 337, 709 (1990)\cr
[26]& D.J. Gross, J.H. Sloane, \NPB 291, 41 (1987)\cr
[27]& R. Gregory, J.A. Harvey, preprint EFI-92-49 (1992)\cr
}
\vfil\eject

\centerline{\bf Figure captions}
\bigskip
{\bf Fig. 1a)} The metric function $R^2$ for $\alpha=1$, $M=10$.

{\bf Fig. 1b)} The metric function $\lambda^2$ for $\alpha=1$, $M=10$.

{\bf Fig. 1c)} The scalar field $\Phi$ for $\alpha=1$, $M=10$.

{\bf Fig. 2a)} The temperature as a function of the mass for several values
of $q$ ($\alpha =1$). The behaviour changes drastically for
$q^2>{73\over 45}$.

{\bf Fig. 2b)} The temperature as a function of the mass for several values of
$q$ ($\alpha=1$). Also in this case the behaviour is very different depending
on $q^2$ being greater, equal or less than $73\over 45$.

\end